\documentstyle[11pt,newpasp,twoside,epsf]{article} 
\def\edcomment#1{\iffalse\marginpar{\raggedright\sl#1\/}\else\relax\fi} 
\reversemarginpar 

\begin{document} 
\title{A new survey of variability in the core of M15 with TRIFFID-2}

\author{Seathr\'{u}n \'{O} Tuairisg, Raymond Butler, Andy Shearer, Michael Redfern}
\affil{National University of Ireland, University Road, Galway, Ireland}
\author{David Butler}
\affil{Max-Planck Institut f\"ur Astronomie, Heidelberg, Germany}

\begin{abstract} 

A two-week campaign of high-resolution imaging of the centre of M15
using the tip/tilt correcting TRIFFID-2 camera, followed by analysis
with ISIS (image matching \& subtraction software), has produced the
most sensitive survey to date of variable stars in the central
$\sim$1$\arcmin$$\times$1$\arcmin$ - a total of 48 were detected -,
and constrains the size of the dwarf nova population.

\end{abstract}

\section{Introduction} 

M15 is a massive core-collapsed cluster, with an enormous stellar
density in the central $\sim$1$\arcmin$$\times$1$\arcmin$, reflected
in the scarcity of photometric variable star detections in this field.
However, Ferraro \& Paresce (1993) used the HST/FOC to identify 19
variable candidates. Our earlier work (Butler et al. 1998), using a
TRIFFID/MAMA camera similar to that described here, resulted in light
curves and periods for all but four of these, with a further 16
suspected new variables.

We made new observations in \emph{B} and \emph{V} simultaneously with
the TRIFFID-2/MAMA camera on the 1m JKT (La Palma), over 12 nights in
July 1997, with seeing of 0$\farcs$7-0$\farcs$8. The 2-d
photon-counted data were first sharpened post-exposure by tip/tilt
correction at 1ms time resolution.  The ISIS (Alard \& Lupton 1998)
image matching and difference imaging package was used to search for
frame-to-frame variability.  All star-like objects 6-sigma above the
local background of the the {\it median} of the set of difference
images were selected as candidate variables (see Figure
1). PSF-fitting photometry was performed at accurate positions for the
candidate variables in the difference images. The resulting relative
fluxes were converted to absolute fluxes by performing profile-fitting
photometery on the reference image. The instrumental magnitudes were
zeropointed against photometry of HST/WFPC2 archival images in the
F555W and F439W filter system. Accurate periods (and hence
lightcurves) were obtained for the set of magnitude measurements for
all candidate variables using the Phase Dispersions Minimisation
technique (PDM, Stellingwerf 1978). Simulations performed using an
artificial image-set modeled on our M15 observations indicate that we
are sensitive to RR Lyrae variables with an amplitude swing of 0.15
mag (at 80\% detection probability). We also increased our sensitivity to dwarf novae (DNe) by summing
all images per night prior to ISIS analysis.

\begin{figure}
\plottwo{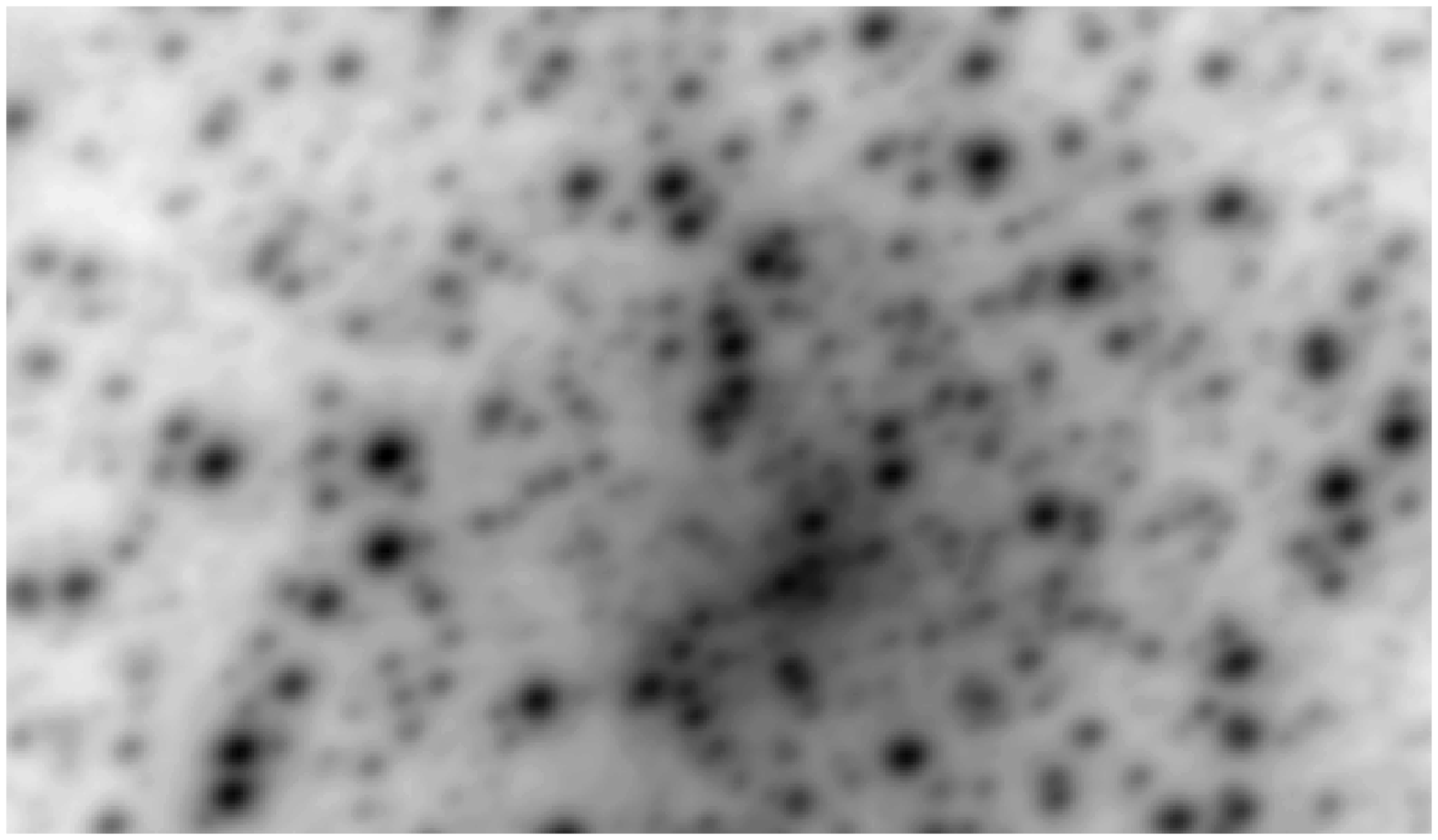}{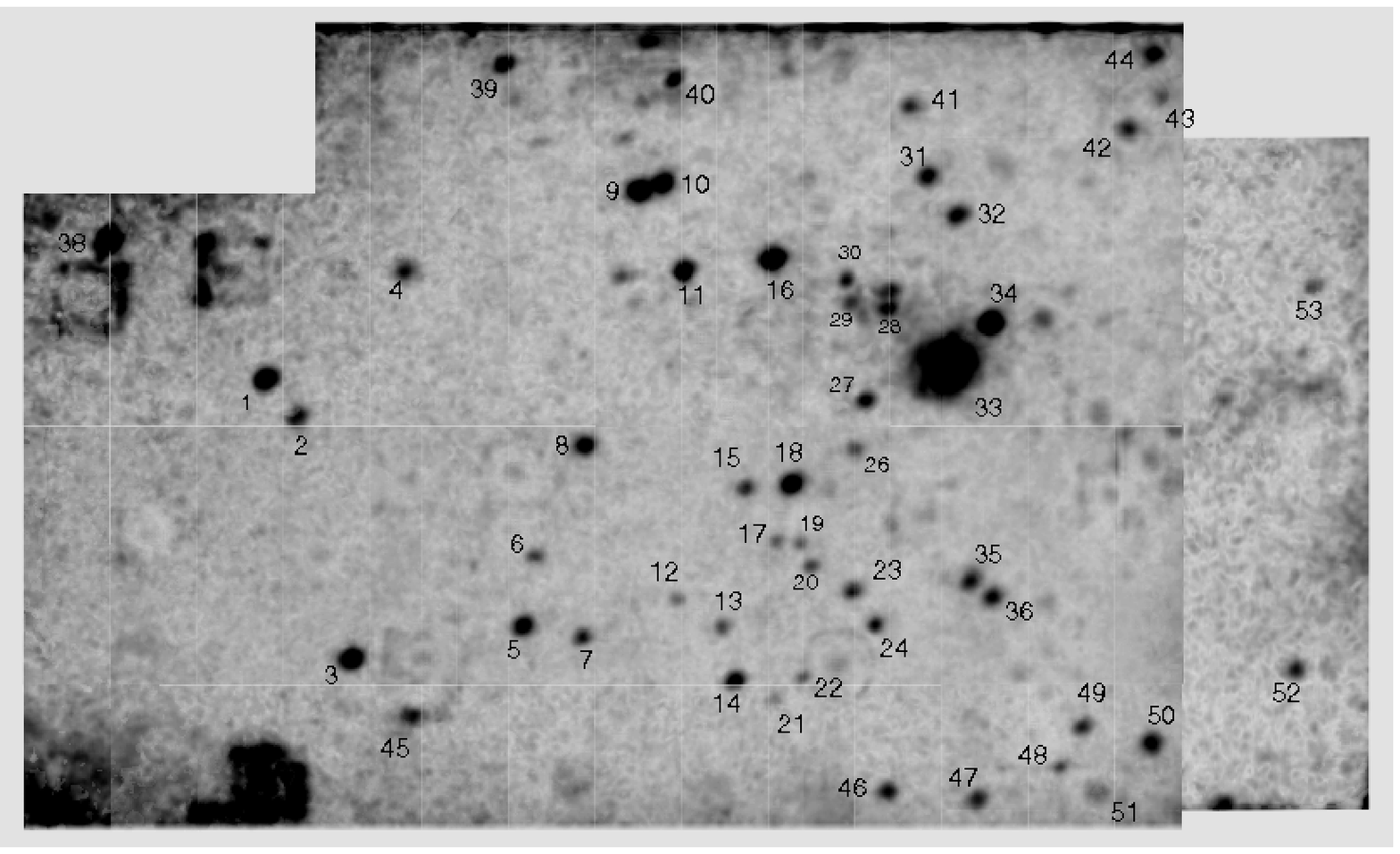}
\caption{Left: one V-band image of M15. Right: median-difference image, clearly showing the variable 
candidates (numbered).}
\end{figure}

\section{Results}

We confirmed all known variables and detected 22 new ones.  The
lightcurve shape, amplitude, scatter and most significant period
distinguished 12 RRab, 19 RRc and 2 (probable) RRd variables.  The
nature of 11 variables remains uncertain. We found that
the core RR Lyrae population has similar periods and amplitudes to the
outer population.  However, in the core we find a marked increase in
the ratio of RRc/RRab stars and a dearth of RRd stars, with
respect to the outer reaches. The latter can be reconciled on
observational grounds but the former cannot. Also detected were the
LMXB counterpart AC211, a W Vir type Cepheid, and a BL Her type
Cepheid - objects 20, 33 and 18 in Figure 1. Our search yielded no
candidate DNe. From simulations, we estimate fewer than 10 (with 92\%
probability) DNe of absolute mag 4.3 (F555W/$V$) to exist in the
studied region. This is the first such quantitative result for M15.

\begin{figure}
\plottwo{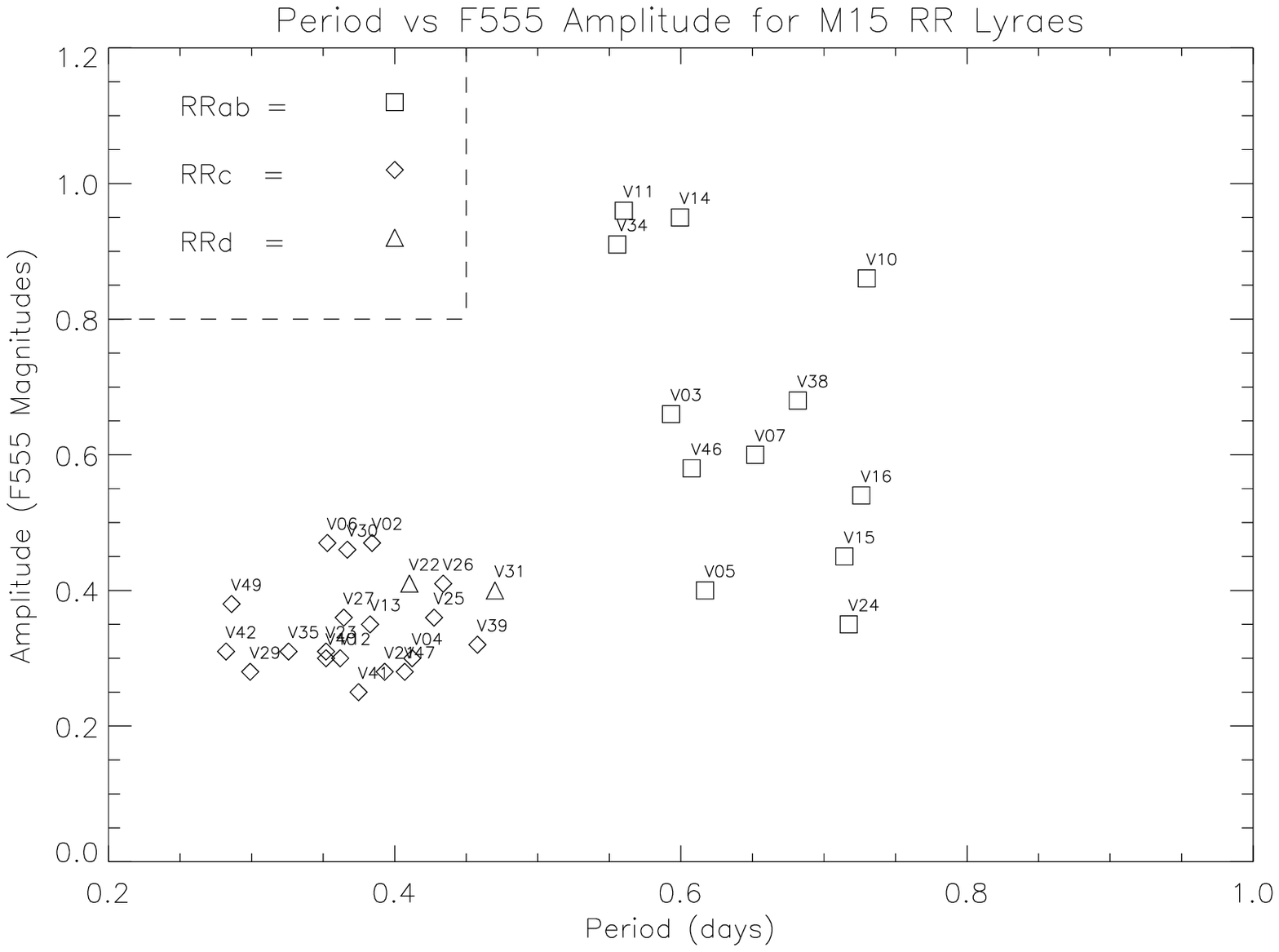}{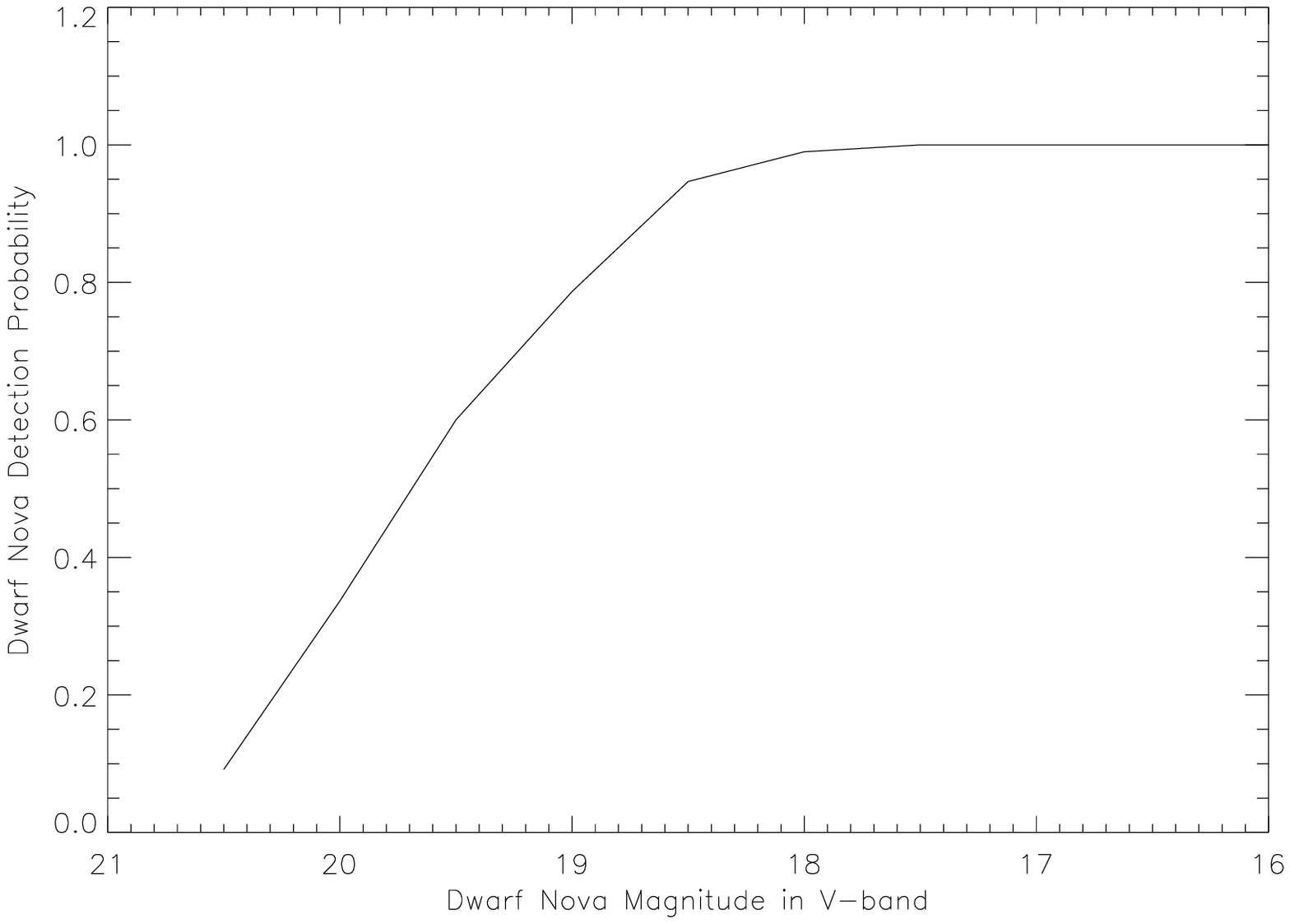}
\caption{Left: RR Lyrae periods vs amplitudes. Right: Probability of detecting a DN in outburst for 3-4 nights versus DN test magnitude.}
\end{figure}

\end{document}